\documentclass[12pt,epsf,amstex]{article}
\usepackage [dvips]{graphicx}
\usepackage{amsmath}
\usepackage{amssymb}
\usepackage{epsfig}

\addtocounter{secnumdepth}{1}
\begin{document}

\newcommand{\br}{\bar{r}}
\newcommand{\bbeta}{\bar{\beta}}
\newcommand{\bgamma}{\bar{\gamma}}
\newcommand{\tbeta}{\tilde{\beta}}
\newcommand{\tgamma}{\tilde{\gamma}}
\newcommand{\bE}{{\bf{E}}}
\newcommand{\bO}{{\bf{O}}}
\newcommand{\bP}{{\bf{P}}}
\newcommand{\bR}{{\bf{R}}}
\newcommand{\bS}{{\bf{S}}}
\newcommand{\bT}{\mbox{\bf T}}
\newcommand{\bt}{\mbox{\bf t}}
\newcommand{\half}{\frac{1}{2}}
\newcommand{\thalf}{\tfrac{1}{2}}
\newcommand{\summ}{\sum_{m=1}^n}
\newcommand{\sumq}{\sum_{q=1}^\infty}
\newcommand{\sumqno}{\sum_{q\neq 0}}
\newcommand{\prodm}{\prod_{m=1}^n}
\newcommand{\prodq}{\prod_{q=1}^\infty}
\newcommand{\maxm}{\max_{1\leq m\leq n}}
\newcommand{\maxphi}{\max_{0\leq\phi\leq 2\pi}}
\newcommand{\tsum}{\Sigma}
\newcommand{\bsA}{\mathbf{A}}
\newcommand{\bsV}{\mathbf{V}}
\newcommand{\bsE}{\mathbf{E}}
\newcommand{\bsT}{\mathbf{T}}
\newcommand{\bsZ}{\hat{\mathbf{Z}}}
\newcommand{\bse}{\mbox{\bf{1}}}
\newcommand{\bspsi}{\hat{\boldsymbol{\psi}}}
\newcommand{\cdottt}{\!\cdot\!}
\newcommand{\deltaR}{\delta\mspace{-1.5mu}R}
\newcommand{\invup}{\rule{0ex}{2ex}}

\newcommand{\bGamma}{\boldmath$\Gamma$\unboldmath}
\newcommand{\dd}{\mbox{d}}
\newcommand{\ee}{\mbox{e}}
\newcommand{\p}{\partial}
\newcommand{\expmVo}{\langle\ee^{-{\mathbb V}}\rangle_0}

\newcommand{\Rav}{R_{\rm av}}
\newcommand{\Rc}{R_{\rm c}}

\newcommand{\la}{\langle}
\newcommand{\ra}{\rangle}
\newcommand{\rao}{\rangle\raisebox{-.5ex}{$\!{}_0$}}  
\newcommand{\rae}{\rangle\raisebox{-.5ex}{$\!{}_1$}}

\newcommand{\beq}{\begin{equation}}
\newcommand{\eeq}{\end{equation}}
\newcommand{\bea}{\begin{eqnarray}}
\newcommand{\eea}{\end{eqnarray}}
\def\lsim{\:\raisebox{-0.5ex}{$\stackrel{\textstyle<}{\sim}$}\:}
\def\gsim{\:\raisebox{-0.5ex}{$\stackrel{\textstyle>}{\sim}$}\:}

\numberwithin{equation}{section}

\thispagestyle{empty}
\title{\Large 
{\bf Heuristic theory for many-faced}\\[2mm]
{\bf $d$-dimensional Poisson-Voronoi cells}\\[2mm] 
\phantom{xxx} }
 
\author{{H.\,J. Hilhorst}\\[5mm]
{\small Laboratoire de Physique Th\'eorique, B\^atiment 210}\\[-1mm] 
{\small Universit\'e Paris-Sud and CNRS}\\[-1mm]
{\small 91405 Orsay Cedex, France}\\}

\maketitle
\begin{small}
\begin{abstract}
\noindent
We consider the $d$-dimensional Poisson-Voronoi tessellation
and investigate the applicability of heuristic methods
developed recently for two dimensions.
Let $p_n(d)$ be the probability that a cell have $n$ neighbors
(be `$n$-faced') and $m_n(d)$ the average 
facedness of a cell adjacent to an $n$-faced cell.
We obtain the leading order terms
of the asymptotic large-$n$ expansions for $p_n(d)$ and $m_n(3)$.
It appears that, just as in dimension two, the Poisson-Voronoi tessellation
violates Aboav's `linear law' also in dimension three. 
A confrontation of this statement with existing Monte Carlo work remains
inconclusive. 
However, simulations upgraded to the level of present-day computer capacity
will in principle be able to confirm (or invalidate) our theory.\\

\end{abstract}
\end{small}
\vspace{45mm}

\noindent LPT Orsay 09/27
\thispagestyle{empty}
\newpage

%%%%%%%%%%%%%%%%%%%%%%%%%%%%%%%%%%%%%%%%%%%%%%%%%%%%%%%%%%%%%%%%%%%%%%%%%%%%%
%%%%%%%%%%%%%%%%%%%%%%%%%%%%%%%%%%%%%%%%%%%%%%%%%%%%%%%%%%%%%%%%%%%%%%%%%%%%%
%%%%%%%%%%%%%%%%%%%%%%%%%%%%%%%%%%%%%%%%%%%%%%%%%%%%%%%%%%%%%%%%%%%%%%%%%%%%%

\section{Introduction} 
\label{secintroduction}

\subsection{General}
\label{secgeneral}

Cellular structures occur in a wide variety of natural systems.
The examples most quoted, but by no means the only ones,
are soap froths and biological tissues. 
Cellular systems also serve as a tool of analysis in a diversity of
problems throughout the sciences and beyond.
Many references may be found in Okabe {\it et al.} \cite{Okabeetal00},
Rivier \cite{Rivier93}, and Hilhorst \cite{Hilhorst05b}.

A popular model of a cellular
structure is the Voronoi tessellation.
It is obtained by performing 
the Voronoi construction on a set of point-like `seeds' in 
$d$-dimensional space. This construction consists of
partitioning space into cells in such a way that 
each point of space is in the cell of the seed to which it is closest.
A $d$-dimensional Voronoi cell is 
convex and is bounded by planar $(d-1)$-dimensional faces.
Two cells are called `neighbors' or `adjacent'
when they have a face in common.
The Voronoi construction may thus serve to define neighbor relations on an
arbitrary set of given seeds.

In the special case that the seed positions are drawn 
randomly from a uniform
distribution, one speaks of the {\it Poisson-}Voronoi tessellation.
It constitutes one of the simplest and best studied
models of a cellular structure.
The analytic study of the statistical
properties of the Poisson-Voronoi tessellation in dimensions $d=2,3$ was
initiated by Meijering \cite{Meijering53} in 1953. 
Monte Carlo results were obtained by several workers in the past decades
(see \cite{Okabeetal00} for references).
The analytical and numerical results
of greatest interest in dimensions $d=2$ and $d=3$
have been listed in Ref.\,\cite{Okabeetal00}.
In arbitrary dimension $d$, a large collection of
statistical properties of the Poisson-Voronoi tessellations were derived in
Refs.\,\cite{Moller89,MollerStoyan07}. 
\vspace{3mm}

One of the most characteristic properties of a cell 
is its number $n$ of faces, 
and it so happens that this is not a property easy to study analytically.
There must be at
least $d+1$ faces, but there may be any number of them.
We denote by $p_n(d)$ the probability that 
a randomly picked cell (all cells with the same probability)
be $n$-faced.
The facedness probability $p_n(d)$ has been the center of interest of
experimental and Monte Carlo work (see \cite{Okabeetal00,Rivier93,Hilhorst05b} 
for references), especially in $d=2$ and $d=3$.
Analytical results for this quantity, however, are very scarce.
Only the one-dimensional case with $p_n(1)=\delta_{n,2}$ is trivial.
Known results in higher dimension include the {\it average\,} facedness 
$\la n \ra_d = \sum_n np_n(d)$ in dimensions $d=2,3,4$.  
Its values are 
$\la n \ra_2=6$, $\la n \ra_3=2+48\pi^2/35=15.535...$, 
and $\la n \ra_4=340/9=37.77...$ \cite{footnotea}. 
It has been determined numerically that
the peaks of the distributions are at $n=6$ for $d=2$
and at $n=15$ for $d=3$; as $n$ increases beyond the peak value, 
$p_n(d)$ decreases very rapidly to zero.  

\subsection{Recent work}
\label{secrecent}

Although in general dimension $d$ 
one readily writes down an $nd$-dimensional integral
for $p_n(d)$, the variables of integration
are coupled in such a way that for all $d>1$ this is a true many-particle
problem. 
As a consequence, it has not been possible to determine the neighbor number
probability $p_n(d)$ analytically.
In dimension $d=2$ progress was made, nevertheless, 
in Refs.\,\cite{Hilhorst05a,Hilhorst05b},
where it was shown that in the limit $n\to\infty$ 
the sidedness probability $p_n(2)$ has the exact asymptotic behavior  
\beq
{p}_n(2) = \frac{C}{4\pi^2}\,\frac{(8\pi^2)^n}{(2n)!} 
\,\big[1+o(1)\big], \qquad n\to\infty,
\label{resultpn2}
\eeq
with $ C = 0.344\,347...$.
The collection of asymptotic results that include (\ref{resultpn2}) 
required considerable calculational effort. Subsequently, however, 
heuristic arguments were developed \cite{Hilhorst09} by which at least part 
of the same results can be derived more easily.
We extend in this work these heuristic methods to
higher dimensions.  
\vspace{3mm}

The correlation between the neighbor numbers of adjacent cells
is usually expressed in terms of the quantity $m_n(d)$, defined as
the average facedness
of a cell that is itself adjacent
to an $n$-faced cell. Aboav's celebrated `linear law' states that 
$nm_n(d)=an+b$ \cite{Aboav70}. It holds within error bars, and with
system-specific $a$ and $b$, for many experimental cellular systems.
In Ref.\,\cite{Hilhorst06} it was demonstrated, however, 
that for the two-dimensional Poisson-Voronoi tessellation Aboav's law 
is in fact a linear approximation limited to small $n$ values.
The true asymptotic behavior appeared to be
\beq
m_n(2)=4+3(\pi/n)^{1/2}+\ldots, \qquad n\to\infty,
\label{resultmn2}
\eeq
and shows that $nm_n(2)$, instead of being linear, has a 
small downward curvature.
In fact, this deviation from the linear law was known from Monte
Carlo simulations \cite{BootsMurdoch83,LeCaerHo90,BrakkeUN}.
In this paper we investigate how (\ref{resultmn2}) is modified 
in dimension $d=3$.

\subsection{This work}
\label{secthiswork} 

This work continues a series of articles 
\cite{Hilhorst05a,Hilhorst05b,Hilhorst06,Hilhorst07,Hilhorst08,Hilhorst09} 
that deal with the properties of Voronoi tessellations.
We consider here the question of what, if anything,
may be learned in higher dimensions
from the two-dimensional case. 
The answer is that the exact methods, as in so many other domains of physics,
remain limited to two dimensions of space.
However, we can apply to higher dimensions the heuristic methods 
that we developed \cite{Hilhorst09} on the basis of the exact ones.

In this way we obtain 
first, in section \ref{secestimatepn},
we obtain a formula for the large-$n$ behavior of $p_n(d)$
in arbitrary space dimension $d$. The argument is a fairly straightforward 
extension from the two-dimensional case \cite{Hilhorst09}.
Then, in section \ref{secaboav},
we obtain a two-term asymptotic large-$n$ expansion
of $m_n$ in dimension $d=3$. 
Our key results are represented by equations 
(\ref{expn3}) and (\ref{resultmn}), that are the higher-dimensional analogs to
(\ref{resultpn2}) and (\ref{resultmn2}), respectively.
One of our findings is that Aboav's law fails also for the 
{\it three}-dimensionsal Poisson-Voronoi tessellation.
In section \ref{seccomparison} we compare our theoretical result for
$m_n(3)$ to existing Monte Carlo work. We conclude in section
\ref{secconclusion}. 

%%%%%%%%%%%%%%%%%%%%%%%%%%%%%%%%%%%%%%%%%%%%%%%%%%%%%%%%%%%%%%%%%%%%%%%%%%%%%
%%%%%%%%%%%%%%%%%%%%%%%%%%%%%%%%%%%%%%%%%%%%%%%%%%%%%%%%%%%%%%%%%%%%%%%%%%%%%

\section{Asymptotic large-$n$ expression for $p_n(d)$}  
\label{secestimatepn}

We consider a $d$-dimensional Poisson-Voronoi tessellation with seed 
density $\rho$.
This density may be scaled to unity but we will
keep it as a check on dimensional coherence.
We select an arbitrary seed, let its position be the origin, and 
are interested in its Voronoi cell, to be referred to as the `central cell'.
The central cell has the same statistical properties as any other cell.
We set ourselves as a first purpose to determine
the facedness probability $p_n(d)$ of this cell in the limit of large $n$.

\subsection{The shell of first neighbor seeds}
\label{secshell}

In the two-dimensional case 
the $n$-sided cell is known for large
$n$ to tend with probability $1$ to a circular shape.
It is natural to assume that in $d$ dimensions the $n$-faced cell
will similarly tend to a hypersphere when $n\to\infty$.
Let $R$ denote the $n$ dependent 
radius of this sphere. Then the central seed has its
$n$ first-neighbor seeds located, for large $n$, 
within a narrow spherical shell of radius $2R$.
We denote the width of this shell by $w$; 
approach to sphericity means
that $w/R\to 0$ as $n\to\infty$.

%%%%%%%%%%%%%%%%%%%%%%%%%%%%%%%%%%%%
%%%%%%%%%%%%%%%%%%%%%%%%%%%%%%%%%%%%
\begin{figure}
\begin{center}
\scalebox{.45}
{\includegraphics{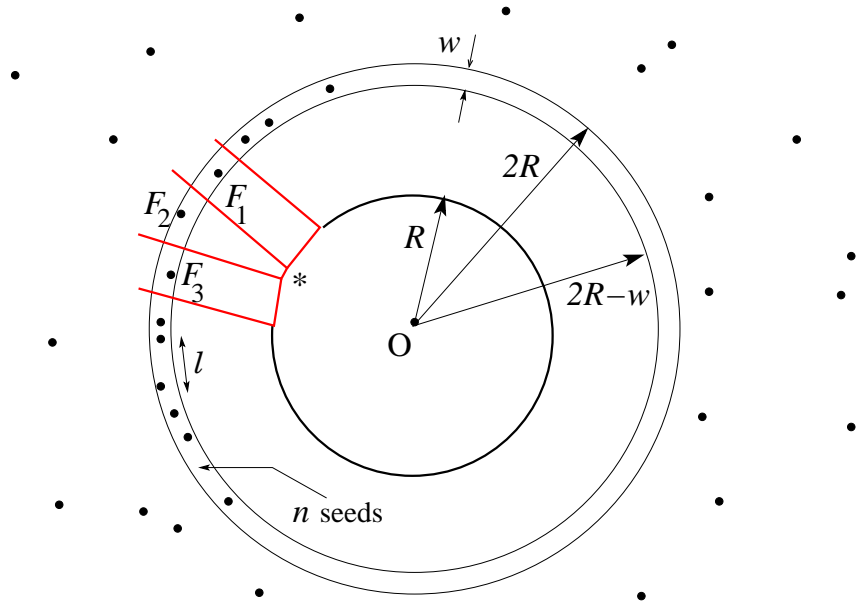}}
\end{center}
\caption{\small 
Schematic representation of the `shell' model of the first-neighbor seeds.
Two hyperspheres (of radii $2R$ and $2R-w$)
are centered about the origin $O$ 
of $d$-dimensional space. A `central' seed is located in $O$.
Other seeds are distributed randomly and uniformly 
with density $\rho$.
The picture shows the rare event of having
(i) $n$ seeds inside the shell;
(ii) no seed inside the inner hypersphere.
The circle of radius $R$ is the approximate boundary
of the central Voronoi cell. 
The asterisk denotes the location of contact between the central cell and the
cell of seed $F_2$.
Further explanation is given in the text.} 
\label{figspheres}
\end{figure}
%%%%%%%%%%%%%%%%%%%%%%%%%%%%%%%%%%%%
%%%%%%%%%%%%%%%%%%%%%%%%%%%%%%%%%%%%

This leads us to the `shell' model 
for the first-order neighbor seeds 
represented in figure \ref{figspheres}. 
We consider two $d$-dimensional hyperspheres
of radii $2R$ and $2R-w$, both centered around the central seed.
Let all seeds other than the central one
be distributed independently and uniformly 
in space with density $\rho$. 
Let now $p_n(d;R,w)$ denote the probability of the
event -- which defines the `shell' model -- that (i) the inner hypersphere 
contains no other seeds than the central one;
and (ii) the shell contains exactly $n$ seeds. 
Our procedure will be
to write $p_n(d;R,w)$ as an explicit function of $n$, $d$, and the
two parameters $R$ and $w$ (this is easy).
We will then, by means of a heuristic argument, 
express $w$ in terms of $R$ and finally maximize 
with respect to $R$. The result will be the desired 
expression for $p_n(d)$.
\vspace{3mm}

The probability $p_n(d;R,w)$ follows from an elementary
calculation and is equal to
\beq
{p}_n(d;R,w)=\frac{(\rho V_{1})^n}{n!}\,\ee^{-\rho V_2},
\label{expn}
\eeq
where $V_1(R,w)$ and $V_2(R)$ are the 
volumes of the shell and of the outer sphere, respectively.
We will denote the volume $v_d$ and the surface area $s_d$ of the
$d$-dimensional hypersphere of unit radius by
\beq
v_d={\pi^{\frac{d}{2}}} / {\left( \tfrac{d}{2} \right)!}\,, \qquad
s_d={2\pi^{\frac{d}{2}}} / {\left( \tfrac{d}{2}-1 \right)!}
\label{exvdsd}
\eeq
Therefore
\bea
V_1(R,w) &=& v_d\left\{ (2R)^d-(2R-w)^d \right\}, \nonumber\\[2mm]
V_2(R) &=& v_d(2R)^d.
\label{defV1V2}
\eea
It will be convenient to work with $\log p_n$. Using (\ref{defV1V2})
in (\ref{expn}) we find
\beq
\log p_n(d;R,w) = -\log n! 
                  - \log\left[ \rho v_d\left\{ (2R)^d-(2R-w)^d \right\} \right]
                  - \rho v_d(2R)^d,
\label{exlogpn}
\eeq
which is exact within the shell model.
This expression for $p_n(d;R,w)$ will be at the basis of what is to follow. 
\vspace{3mm}

%%%%%%%%%%%%%%%%%%%%%%%%%%%%%%%%%%%%
%%%%%%%%%%%%%%%%%%%%%%%%%%%%%%%%%%%%
\begin{figure}
\begin{center}
\scalebox{.60}
{\includegraphics{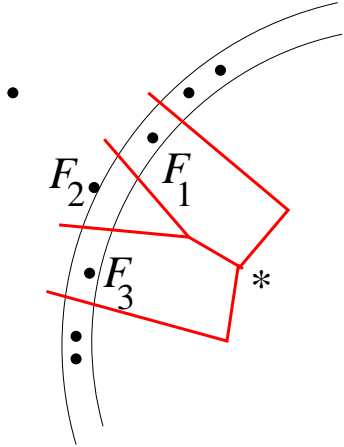}}
\end{center}
\caption{\small Detail of figure \ref{figspheres} in which seed $F_2$ has
  moved a little outward and is now screened by its neighbors $F_1$ and $F_3$
  from being in contact with the central cell.} 
\label{figscreening}
\end{figure}
%%%%%%%%%%%%%%%%%%%%%%%%%%%%%%%%%%%%
%%%%%%%%%%%%%%%%%%%%%%%%%%%%%%%%%%%%

%%%%%%%%%%%%%%%%%%%%%%%%%%%%%%%%%%%%
%%%%%%%%%%%%%%%%%%%%%%%%%%%%%%%%%%%%
\begin{figure}
\begin{center}
\scalebox{.45}
{\includegraphics{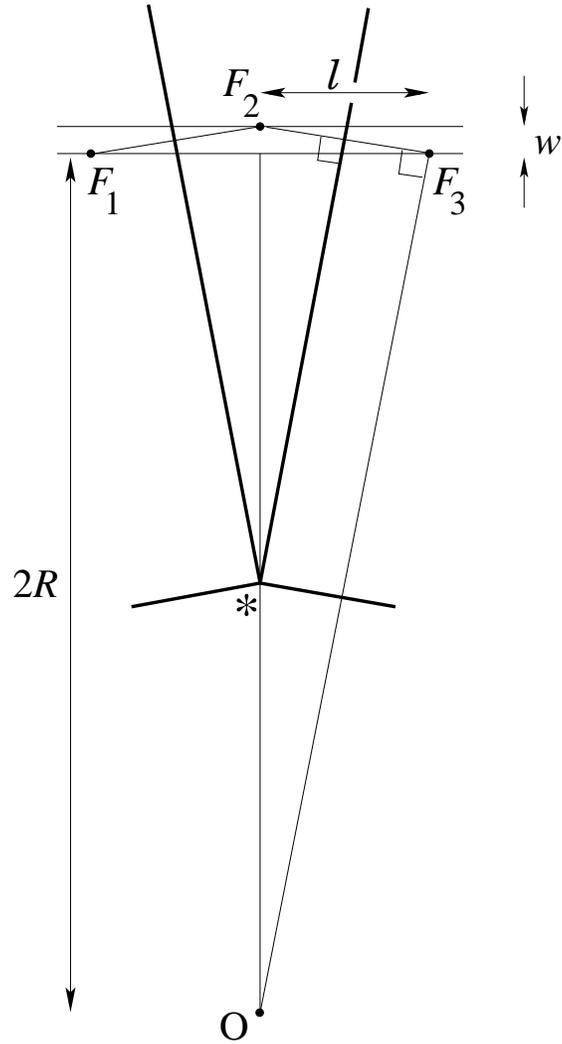}}
\end{center}
\caption{\small Marginal situation in between those of figures
\ref{figspheres} and \ref{figscreening}. The central cell has a point
contact with the cell of seed $F_2$ in the four-vertex denoted by the
asterisk.} 
\label{figwlR}
\end{figure}
%%%%%%%%%%%%%%%%%%%%%%%%%%%%%%%%%%%%
%%%%%%%%%%%%%%%%%%%%%%%%%%%%%%%%%%%%

\subsection{Relation between $w$ and $R$}

We now look for a relation between $w$ and $R$.
Figure \ref{figspheres} is a schematic two-dimensional representation of 
a $d$-dimensional situation. It shows part of
the Voronoi cell boundaries of three first-order
neighbor seeds $F_1$, $F_2$, and $F_3$.
Each line segment in the figure belongs to a $(d-1)$-dimensional face
that perpendicularly bisects the line segment joining two neighboring seeds. 
An asterisk marks the short line segment 
common to the central cell and the cell of $F_2$.
As shown in figure \ref{figscreening},
this face is so small that it disappears when $F_2$, roughly speaking, crosses
to the outside of the outer hypersphere:
$F_2$ is then be `screened' by its neighboring seeds $F_1$ and $F_3$,
When $w$ is small with respect to the typical
interseed distance $\ell$ in the shell, then such screening will be negligible
for a random arrangement of the $n$ seeds in the shell.
We now relate shell width $w$ to the radius 
$R$ by imposing that $w$ be
small enough for this to be the case,
but otherwise as large as possible.
Figure \ref{figwlR} depicts a
marginal situation in which the screening of
$F_2$ by $F_1$ and $F_3$ sets in.
The idealization consisting in placing $F_1$ and $F_3$ at equal distances from
the origin and symmetrically with respect to $F_2$ is good enough for our
ourpose. Elementary geometry then shows that
$w$, $\ell$, and $R$ are related by
\beq
\frac{w}{\ell} \sim\frac{\ell}{2R}\,.
\label{relwellR}
\eeq
where the symbol\, $\sim$\, denotes proportionality in the limit of large $n$.
We now wish to eliminate $\ell$ from this relation.
\vspace{3mm}

The shell may be considered locally as a flat $(d-1)$-dimensional space if 
\beq
w \ll \ell \ll R.
\label{ineqwellR}
\eeq
We will asume here, and be able to verify afterwards, that in the limit
of large $n$ these conditions are satisfied.
The typical interseed distance $\ell$ between the
first-neighbor seeds in the shell is then easily found.
For $w \ll \ell$ the $n$ seeds may be considered as distributed 
on a $(d-1)$-dimensional hypersurface of area 
\beq
S = s_{d}(2R)^{d-1}. 
\label{exsurface}
\eeq
They therefore have a $(d-1)$-dimensional
surface density $\sigma=n/S$,
whence it follows that their typical distance $\ell$ may be defined by
\beq
v_{d-1}\ell^{d-1}=Sn^{-1}.
\label{eqnell}
\eeq
From (\ref{exsurface}) and (\ref{eqnell}) it follows that
\beq
\ell = 2R\left( \frac{s_{d}}{nv_{d-1}} \right)^{ \frac{1}{d-1} }.
\label{solell}
\eeq 
Comparing finally (\ref{solell}) and (\ref{relwellR}) 
suggests that in the shell
model we should set 
\bea
w = 2R (c_d n)^{ -\frac{2}{d-1} },
\label{relwR}
\eea
where $c_d$ is a (not exactly known)
numerical constant of order unity. 
Whereas have argued above for the validity of (\ref{relwR}) in the limit of
asymptotically large $n$ and $R$, we adopt it now as part of the
definition of the shell model for arbitrary $w$ and $R$.
At this point it may be verified that the necessary 
conditions (\ref{ineqwellR}) both hold if
\beq
(c_d n)^{-\frac{1}{d-1} } \ll 1.
\eeq
Equation (\ref{relwR}) is the desired relation between $w$ and $R$.

\subsection{Maximizing the entropy}
\label{secmaxent}

Upon substituting (\ref{relwR}) in (\ref{exlogpn})
and denoting the result by $p_n(d;R)$ we obtain
\beq
\log p_n(d;R) = -\log n! + n\log\rho v_d(2R)^d - \rho v_d(2R)^d 
+ n\log Z_n(d)
\label{exlogpn2}
\eeq
where
\vspace{-5mm}

\bea
Z_n(d) &=& 1-\left\{ 1-(c_dn)^{-\frac{2}{d-1} } \right\}^d 
\nonumber\\[2mm]
       &\simeq& d(c_dn)^{-\frac{2}{d-1}},
\label{defZnd}
\eea
in which the second line represents the leading order 
behavior as $n\to\infty$.
\vspace{3mm}

Equation (\ref{exlogpn2}) represents the entropy of the arrangement of seeds
and still contains $R$ as a free parameter. 
It is again easy to maximize $\log p_n(d;R)$ expression 
with respect to $R$.

Upon varying the right hand side of (\ref{exlogpn2}) with respect to $R$ 
we find that it has its maximum for $R=R_*$  where 
\beq
2R_* = \left( \frac{n}{\rho v_d} \right)^{\frac{1}{d}}. 
\label{exRstar}
\eeq
The corresponding $w_*$ follows from substitution of 
(\ref{exRstar}) in (\ref{relwR}).
We are now ready to obtain our heuristic result as the probability that
maximizes the configurational entropy, that is, $p_n(d)=p_n(d;R_*)$. 
Substitution of (\ref{exRstar}) and (\ref{defZnd})
in (\ref{exlogpn2}) yields
\beq
\log p_n(d) = -\log n! + n\log n -n +n\log\left[d(c_dn)^{-\frac{2}{d-1}}\right]
           + n\log z_n(d),
\label{exlogpn3}
\eeq
in which
\bea
\epsilon_n(d) &=& d^{-1}(c_dn)^{\frac{2}{d-1}}Z_n(d)-1 \nonumber\\[2mm]
&=& \sum_{k=1}^{d-1}\frac{(-1)^k}{d}\binom{d}{k+1}
         \left(c_dn\right)^{-\frac{2k}{d-1}}
\label{exznd}
\eea
is a polynomial in $(c_dn)^{-\frac{2}{d-1}}$ without constant term.
This may be rearranged to yield the large-$n$ expansion
\beq
\log p_n(d) = -\log\left[\left(\tfrac{2n}{d-1}\right)!\right]
  \,+\, n\log A_d
  + n\log\left[1 + \epsilon_n(d)\right] 
  + \tfrac{1}{2}\log\tfrac{2}{d-1} + o(1),
\label{exlogpn4}
\eeq
where $A_d$ is an abbreviation for
\beq
A_d = d\left[\tfrac{1}{2}(d-1)c_d\,\ee\right]^{-\frac{2}{d-1}}.
\label{defalphad}
\eeq
Hence we may write
\beq
p_n(d)\,=\,\left(\tfrac{2}{d-1}\right)^{\frac{1}{2}}\,
\frac{A_d^n}{\left(\frac{2n}{d-1}\right)!}
\left[\,1\,+\,\epsilon_n(d)\right]^n\,\left[\,1\,+\,o(1)\right],
\qquad n\to\infty,
\label{expnd}
\eeq
which is our final result for general $d$.

Since $1+\epsilon_n(d)$ has to be elevated to the
$n$th power, this polynomial cannot be included with the $o(1)$ terms
when $d>3$.
Dimension $d=3$ is a marginal case and $\epsilon_n(3)$ contributes
to the constant prefactor of $p_n(3)$. Hence $p_n(3)$ is given by
\beq
p_n(3)\,=\,C\,\frac{A_3^n}{n!}\,\left[\,1\,+\,o(1)\right],
\qquad n\to\infty,
\label{expn3}
\eeq
in which $C=\exp(-1/c_3)$ and $A_3=1/(3c_3\ee)$ are numerical constants.

\subsection{Comments}
\label{seccomments}

The arguments of this section are heuristic. Their validity is best assessed
by a comparison to the two-dimensional case where exact results are available.
These strongly suggest that 
the inverse factorial $1/\left( \frac{2n}{d-1} \right)!$, which is
the dominant factor in the
large-$n$ behavior in (\ref{expnd}) and (\ref{expn3}), are exact. 
They also lead us to believe that the functional form of $p_n(d)$,
that is, an exponential divided by a factorial, can be trusted.
The value of the numerical constant $c_d$ and hence of $A_d$ remain, however, 
undetermined, since $c_d$ appears in the theory only as a proportionality
constant in the order-of-magnitude estimate (\ref{relwR}). 
Finally,
the decay of $p_n(d)$ with growing $n$ is less fast than in dimension $d=2$,
where $p_n(2)\sim A_2^n/(2n)!$
as shown by (\ref{resultpn2}).

%%%%%%%%%%%%%%%%%%%%%%%%%%%%%%%%%%%%
%%%%%%%%%%%%%%%%%%%%%%%%%%%%%%%%%%%%
\begin{figure}
\begin{center}
\scalebox{.45}
{\includegraphics{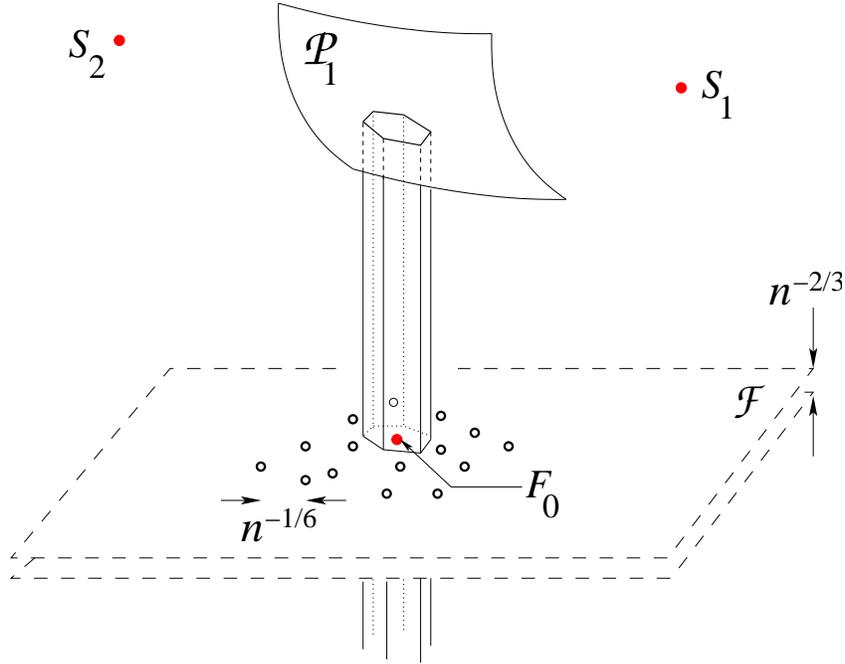}}
\end{center}
\caption{\small The plane ${\cal F}$,
whose actual width $w_*\sim n^{-2/3}$ may be set to zero,
contains the seeds that are
first-neighbors to the central seed (the latter is located
at a distance $2R_*\sim n^{1/3}$ below the plane and is not shown).
$S_1$ and $S_2$ are second-neighbor seeds. ${\cal P}_1$ is the paraboloid of
all points equidistant from $S_1$ and from ${\cal F}$. 
The `prism' is the Voronoi cell of an arbitrarily selected first-neigbor
seed $F_0$. Each vertical face of this
prism lies in a plane equidistant from a pair of points in ${\cal F}$.
The intersection of the prism with ${\cal F}$ 
is the boundary of the two-dimensional
Voronoi cell of $F_0$ in ${\cal F}$.
The prism intersects the parabola ${\cal P}_1$ according to a polygon
which is the projection of this boundary.
It defines the face that the Voronoi cells of $F_0$ and $S_1$ have in common.
In the downward direction the Voronoi cell of $F_0$ ends by a face (not shown)
in common with the central cell.}
\label{fig1}
\end{figure}
%%%%%%%%%%%%%%%%%%%%%%%%%%%%%%%%%%%%
%%%%%%%%%%%%%%%%%%%%%%%%%%%%%%%%%%%%

%%%%%%%%%%%%%%%%%%%%%%%%%%%%%%%%%%%%
%%%%%%%%%%%%%%%%%%%%%%%%%%%%%%%%%%%%
\begin{figure}
\begin{center}
\scalebox{.45}
{\includegraphics{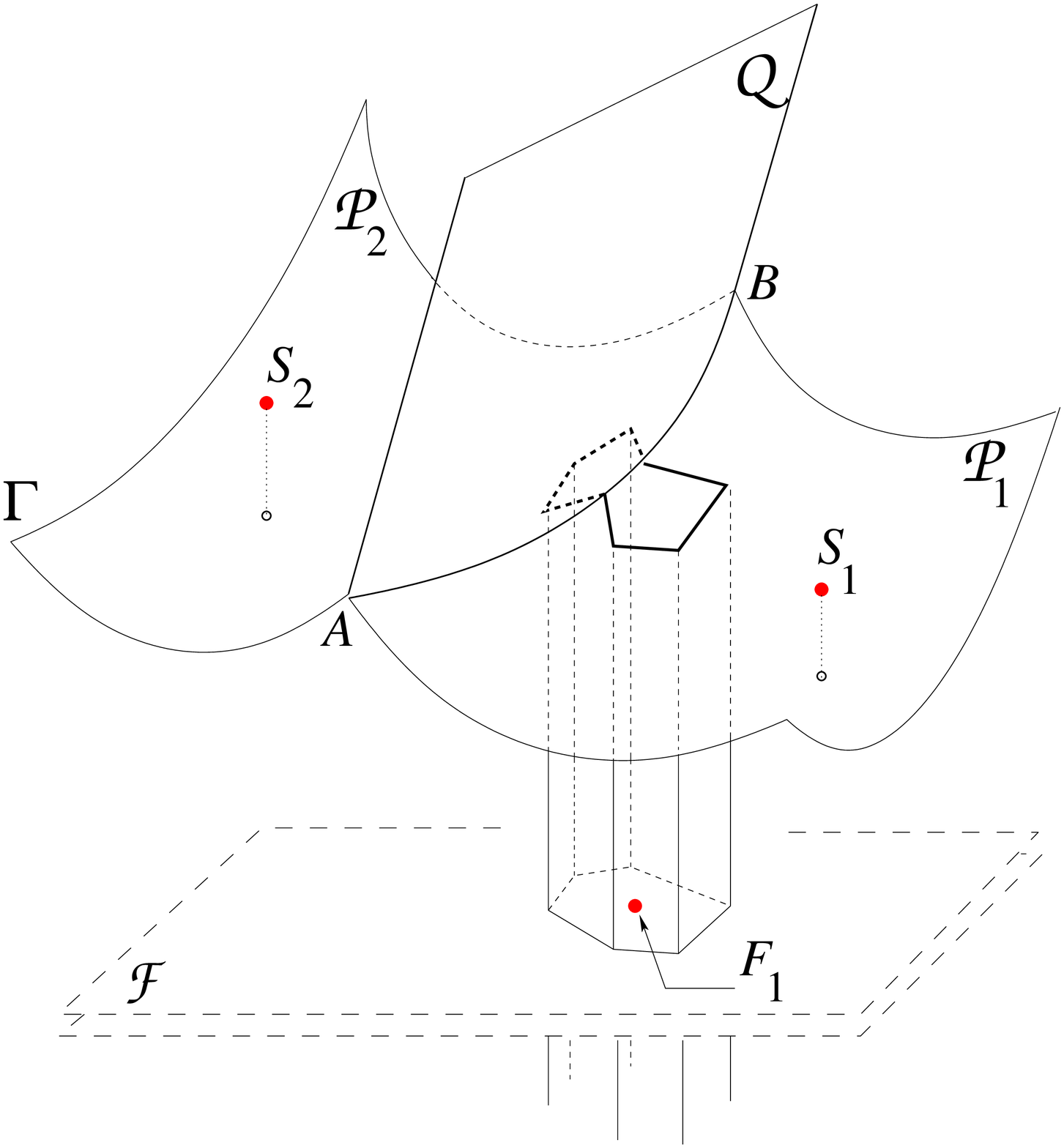}}
\end{center}
\caption{\small The piecewise paraboloidal surface $\Gamma$ is the locus of
points equidistant from the plane of first-neighbor seeds, ${\cal F}$,
and the set of second-neighbor seeds $\{S_i\}$
constitutes a piecewise paraboloidal surface called $\Gamma$. All contributing
paraboloids are circular and have their axes perpendicular to ${\cal F}$. 
In particular,
${\cal P}_1$ and ${\cal P}_2$ are the paraboloids of points equidistant
from the seeds $S_1$ and $S_2$, respectively, and the plane ${\cal F}$.
The plane ${\cal Q}$ bisects the line segment $S_1S_2$ perpendicularly.
The curved segment  $AB$ lies on the parabola along which
${\cal P}_1$, ${\cal P}_2$, and ${\cal Q}$ intersect. 
The Voronoi cell of the first-neighbor seed $F_1$ happens to intersect
$\Gamma$ at the intersection $AB$ between two paraboloids. 
As a result this Voronoi cell has not one, but two faces at its upper end.} 
\label{fig2}
\end{figure}
%%%%%%%%%%%%%%%%%%%%%%%%%%%%%%%%%%%%
%%%%%%%%%%%%%%%%%%%%%%%%%%%%%%%%%%%%

%%%%%%%%%%%%%%%%%%%%%%%%%%%%%%%%%%%%%%%%%%%%%%%%%%%%%%%%%%%%%%%%%%%%%%%%%%%%%

\section{Aboav's law in dimension $d=3$}
\label{secaboav}

\subsection{The plane ${\cal F}$ of first neighbors}
\label{secplaneF}

For large $n$ the first-neighbor seeds are arranged 
in a nearly spherical shell. 
Its radius, according to (\ref{exRstar}) and (\ref{exvdsd}), is equal to
 $2R_*=\frac{1}{4}(6n/\pi\rho)^{1/3}$ and its width,
according to (\ref{exRstar}) and (\ref{relwR}), is 
$w_* \sim (\rho n^2)^{-1/3}$. 
This width may be set to zero for all considerations of this section,
which means neglecting the small random deviations of the radial coordinates
of the first neighbors.
The surface area of the sphere being 
$S=4\pi(2R_*)^2=\pi^{1/3}(6n/\rho)^{2/3}$, 
the typical interseed distance between the first-neigbor seeds is 
$\ell\sim(S/n)^{1/2}\sim (\rho^2 n)^{-1/6}$.
On the scale of the interseed distances
we may therefore consider the shell as a flat surface
that we will denote by ${\cal F}$ and also refer to as a `plane'.
This situation has been represented in figure \ref{fig1}.

In the limit of large $n$ the first-neighbor cells 
become very elongated prism-like objects, as already begins to be apparent in 
figure \ref{figspheres} (snapshots of realistic {\it two}-dimensional 
many-sided cells with $n$ as high as $1500$ are shown 
in reference \cite{Hilhorst07}).
Each first-neighbor cell has one face in common with the central cell.
With the width of ${\cal F}$ set to zero,
the faces between adjacent first neighbors are
perpendicular to ${\cal F}$ and define the sides of
a prism around each first meighbor.
These prisms intersect the plane ${\cal F}$ according to a 
two-dimensional cellular structure. 
The typical cell area in ${\cal F}$ is $a=S/n \sim \ell^2
\sim (\rho^2 n)^{-1/3}$.
The seeds are not uniformly (Poisson) distributed in ${\cal F}$ but will
effectively repel each other; nevertheless,
just for topological reasons, the cells in the plane ${\cal F}$ have an
average of exactly six neighbors.

\subsection{First and second neighbors}
\label{secfirstsecond}

We consider now the faces between the first and second-neighbor cells.
The second-neighbor seeds are marked $S_1, S_2,\ldots$ in figures
\ref{fig1} and \ref{fig2}. There is no restriction on their positions 
as long as they stay out of the sphere of radius $2R_*$,
and the typical distance $\rho^{-1/3}$ between them is independent of $n$.
We denote by $\Gamma$ the surface of points that are equidistant from
${\cal F}$ and from the set $\{S_i\}$ of second neighbors.
Since for $n\to\infty$ the first-neighbor seeds become infinitely dense 
in ${\cal F}$, in that limit $\Gamma$ is a piecewise paraboloidal surface.
The paraboloids join along lines of intersection (`seams') that are 
segments of parabolas. For example, the curve $AB$ in figure \ref{fig2} 
lies on such a parabolic seam. 
For the considerations that follow
it will be convenient to project $\Gamma$ onto ${\cal F}$. 
The set of parabolic seams of $\Gamma$ will project onto 
${\cal F}$ as a two-dimensional cellular net of trivalent vertices, 
connected by segments of parabolas.
We will refer to the cells of this network as `supercells' in order to
distinguish them from the `ordinary' cells (discussed above) 
due to the intersections of the prisms with ${\cal F}$.
The typical supercell area will be of order $\sim n^0$ as $n\to\infty$.
Since the radius of the spherical shell behaves as
$R_*\sim n^{1/3}$, it is well approximated by
the flat surface ${\cal F}$ also at the scale of the supercells.

We analyze now, within the plane ${\cal F}$, the intersection of the net 
of supercells with the ordinary cells. 
In the limit $n\to\infty$ the fraction
of ordinary cells {\it not\,} intersected by a segment of the supercell net 
will tend to unity. For reasons that will become clear just below
we denote this fraction of ordinary cells by $f_8$. 
The prism (three-dimensional cell) that encloses a cell of this type, 
will therefore be bounded below by the central cell and 
above by a {\it single\,} second-neighbor cell.
Since (for mere topological reasons)
such cells are adjacent to, on average, six other first-neighbor cells, 
their total number of neighbors is eight.
 
There are, however, two special types of ordinary cells: 
(a) those intersected by a perimeter segment of a supercell; and 
(b) those containing the vertex where three such perimeter segments join.
In figure \ref{fig2} the cell of seed $F_1$ is an example of
a special cell of type (a).
These two special types of cells will represent fractions
of all ordinary cells to be denoted $f_{9}$ and $f_{10}$, respectively.
A counting similar to the one above easily shows that the corresponding types
of three-diemensional cells have $9$ and $10$ neighbors, respectively. 

For $n\to\infty$ the fractions $f_9$ and $f_{10}$ will vanish, 
and we will now determine exactly how.
Since the supercells are of linear dimension $\sim n^0$
and the ordinary cells of linear dimension $\sim n^{-1/6}$,
a supercell will contain $\sim n^{1/3}$ ordinary cells.
Only a finite number of these (on average six) will be located on the vertices
of the supercell, and therefore we deduce that $f_{10}=a_2n^{-1/3}$ as
$n\to\infty$, where $a_2$ is a numerical constant.
The parabolic segments of a supercell perimeter 
are straight lines at the scale of the ordinary cells.
Typically such a perimeter segment will therefore intersect $\sim n^{1/6}$
ordinary cells. Hence $f_9$ is of order $n^{-1/6}$. Assuming an expansion in
powers of $n^{-1/6}$ we will write $f_9=a_1n^{-1/6}+b_1n^{-1/3}$,
where $a_1$ and $b_1$ are numerical constants.

To order $n^{-1/3}$ we have that $f_8+f_9+f_{10}=1$.
Hence to this order
the average number of neighbors $m_n(3)$ of a cell with $n$ neighbors is
given by
\beq
m_n(3) = 8f_8+9f_9+10f_{10}\,.
\label{exprmn}
\eeq
Substituting the above expressions for the $f_n$\, yields
\beq
m_n(3)\,=\,8\,+\,k_1n^{-1/6}\,+\,k_2n^{-1/3}\,+\ldots\,,
\label{resultmn}
\eeq
in which $k_1=a_1$ and $k_2=b_1+2a_1$.
The constants $k_1>0$ and $k_2$ are unknown. 
Equation (\ref{resultmn}) is the two-dimensional counterpart of
(\ref{resultmn2}). It shows that in dimension $d=3$ 
Aboav's linear law {\it cannot hold\,} for $n$ asymptotically large.  
This law therefore is necessarily an approximation (and possibly a very good
one) in the experimentally accessible window of $n$ values.
Below we will study the deviation from Aboav's law numerically.

\section{Comparison to simulation data}
\label{seccomparison}

\begin{figure}
\begin{center}
\scalebox{.45}
{\includegraphics{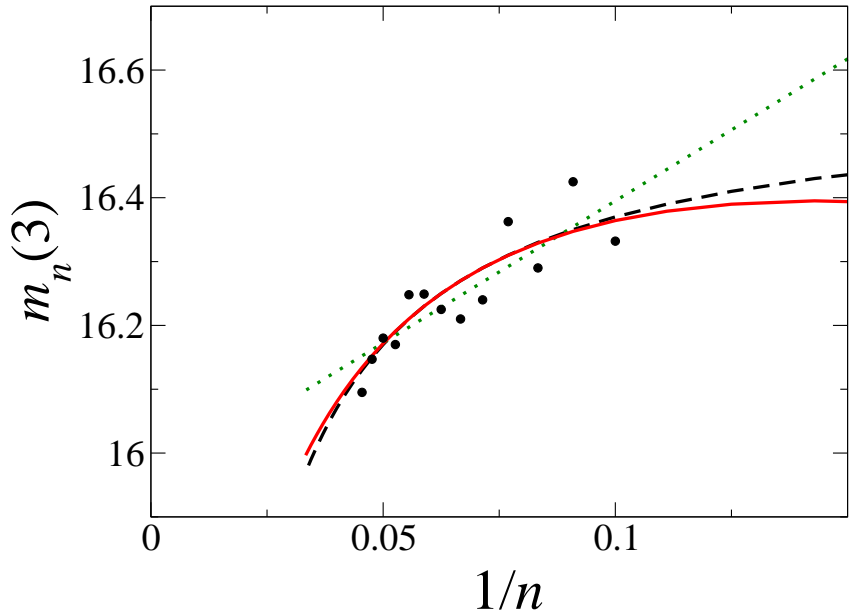}}
\end{center}
\caption{\small 
Dots: data points by Kumar {\it et al.} \cite{Kumaretal92}.
Straight dotted line: Aboav's law with parameters as in
(\ref{Fortesfit}) due to Fortes \cite{Fortes93}. 
Dashed curve : nonasymptotic fit (\ref{Kumarfit}) due to Kumar {\it et al.}
Solid curve: our theoretical equation (\ref{resultmn}) 
for the values of $k_1$ and $k_2$ given in the text.} 
\label{figmn}
\end{figure}

In the preceding sections we derived results that are asymptotic in $n$,
whereas data are mostly in a certain range of small $n$.
In earlier work it turned out \cite{Hilhorst06}, however,
that the asymptotic expression for the
nearest-neighbor correlation $m_n(2)$  gives a good approximation 
to the two-dimensional data for {\it all} $n$ values.
We may therefore hope that equation (\ref{resultmn})
will similarly provide a good fit to the three-dimensional data.
Only few such data exist.
The ones most relevant are due to Kumar {\it et al.} \cite{Kumaretal92}.
These authors determined by simulation, among several other quantities,
the probability $p_n(3)$ and 
the correlation $m_n(3)$ in the range $10\leq n\leq 22$.
Their $p_n(3)$ data, not shown here, peak at $n=15$.
We have reproduced their $m_n(3)$ data in our figure \ref{figmn}.
Following Earnshaw and Robinson
\cite{EarnshawRobinson94}, we plot $m_n$
as a function of $1/n$ (rather than $m_n$ or $nm_n$ as a function of $n$). 
In the $m_n$ {\it versus\,} $1/n$ plot Aboav's law again
corresponds to a straight line, but deviations from linearity are easier to
detect.

Within the measurement window
$m_n(3)$ is clearly seen to decrease 
with $n$, but only from around $16.4$ to $16.1$. 
Kumar {\it et al.} fitted this behavior by
\beq
m_n(3)=16.57-0.02n,
\label{Kumarfit}
\eeq
represented by the dashed line in figure \ref{figmn}. This relation obviously
cannot be asymptotic.
In a later analysis of the same data, Fortes \cite{Fortes93} proposed to fit
them by Aboav's law, namely
\beq
m_n(3)=15.95+\frac{4.45}{n}\,.
\label{Fortesfit}
\eeq
It is shown as the straight dotted line in figure \ref{figmn}.

%%%%%%%%%%%%%%%%%%%%%%%%%%%%%%%%%%%%
%%%%%%%%%%%%%%%%%%%%%%%%%%%%%%%%%%%%
%%%%%%%%%%%%%%%%%%%%%%%%%%%%%%%%%%%%
%%%%%%%%%%%%%%%%%%%%%%%%%%%%%%%%%%%%

We wish to compare these two earlier fits
to our theoretical functional form,
equation (\ref{resultmn}). To that end we choose the constants 
$k_1$ and $k_2$ such that in $n=15$ (where $p_n$ is maximum)
our curve and the dashed fit produce the same values of $m_n(3)$
and its $n$-derivative,
that is, $m_{15}(3)=16.27$ and $m'_{15}(3)=-0.02$.
This leads to $k_1=23.15$ and $k_2=-15.96$.
The result is the solid curve shown in the figure. 
We emphasize that this procedure involves the additional
assumption that it is correct in the finite-$n$ regime
to use the asymptotic expression
(\ref{resultmn}) with all terms beyond order $n^{-1/3}$ discarded.

Returning now to figure \ref{figmn}, we observe that
the simulation data scatter too much to be able to unambiguously
distinguish between the three curves.
%The downward curvature of $nm_n(3)$ implied by our functional form 
%(\ref{resultmn}) appears to be invisible to the eye in the plot of 
%figure \ref{figmn}.
%In Ref.\,\cite{Hilhorst06} the ratio $\kappa=-(nm_n)^{''}/(nm_n)$, 
%evaluated in the maximum of the distribution $p_n$, 
%was taken as a measure of this curvature; if Aboav's law were
%valid, it would vanish. 
%The earlier result, $\kappa=0.004$ for dimension $d=2$
%\cite{Hilhorst06}, may now be compared to the present one,
%$\kappa=0.000\,17$ for $d=3$. 
%There is no experimental system that we know of in which such a small effect
%would be observable.
The considerations of this section point to a most interesting question:
can one establish by
Monte Carlo simulation the presence of the downward curvature in the
$m_n$ {\it versus\,} $1/n$ plot in three dimensions?
Curvature, although not proving (\ref{resultmn}), would at least rule out
Aboav's law.
Simulations at least an order of magnitude larger than the existing ones
will be necessary; this however is within present-day 
machine capacity.

%%%%%%%%%%%%%%%%%%%%%%%%%%%%%%%%%%%%%%%%%%%%%%%%%%%%%%%%%%%%%%%%%%%%%%%%%%%

\section{Conclusion}
\label{secconclusion}

We have considered
Poisson-Voronoi diagrams in spatial dimensions $d$ higher than two. 
We obtained analytic expressions 
for (i) the facedness (or : neighbor number) probability $p_n(d)$ 
and (ii) the two-cell correlation $m_n(3)$,
both valid in the limit of asymptotically large $n$. 
We conclude that Aboav's law cannot be strictly valid in dimension $d=3$,
although it may be a very good approximation in the regime most easily
accessible to experiment and simulation.
These results rest on heuristic arguments developed in analogy 
to reasoning previously shown \cite{Hilhorst09} to be valid 
in two dimensions. 
They cannot be considered as mathematically proved,
but their value derives from the fact that
this is the only theoretical work so far in this direction.
We believe that confirmation of the failure of Aboav's law
in three dimensions is within the reach of Monte Carlo simulations
that are possible today.

%%%%%%%%%%%%%%%%%%%%%%%%%%%%%%%%%%%%%%%%%%%%%%%%%%%%%%%%%%%%%%%%%%%%%%%%%%%

%\section*{Acknowledgments}

%%%%%%%%%%%%%%%%%%%%%%%%%%%%%%%%%%%%%%%%%%%%%%%%%%%%%%%%%%%%%%%%%%%%%%%%%%%

\appendix % ?


\begin{thebibliography}{10}
\bibitem{footnotea} %7
The value of $\la n\ra_4$ may be deduced from the relations given in
Ref.\,\cite{Moller89}.
\end{thebibliography}
\end{document}